\documentstyle[12pt]{article}
\newcommand{\h}{\hspace{0.6in}}
\newcommand{\nh}{\hspace{0.2in}}

\begin{document}
\title{ EXPANSION ASPECT OF COLOR TRANSPARENCY ON THE LATTICE }
\author{D.~Makovoz and G.~A.~Miller}
\maketitle
{\em Department of
Physics, FM-15, University of Washington, Seattle, WA 98195,   USA}\\
\vspace{1.0cm}\\
\begin{abstract}
The opportunity to observe color transparency (CT) is determined by
how rapidly a small-sized hadronic wave packet expands.
Here we use
SU(2) lattice gauge theory with Wilson fermions in the quenched
 approximation to investigate the expansion.
The wave packet is modeled by
a point hadronic source, often used as an
interpolating field in lattice calculations. The procedure is to
determine the Euclidean time (t), pion channel,
Bethe-Salpeter amplitude  $\Psi(r,t)$, and then
evaluate
  $b^2(t)=\int d^3 r \Psi(r,t) r^2 sin^2 \theta \Psi_{\pi}(r)$.
 This quantity  represents the soft interaction
of a small-sized wave packet with a pion.
The time dependence of
$b^2(t)$ is fit as a superposition of three states, which
is found sufficient to reproduce
a reduced size wave packet.
Using this superposition allows us
to make the analytic continuation required
to study the wave packet expansion in real time.
We find that the matrix elements of the soft interaction $\hat b^2$ between
the excited and ground state decrease rapidly with the energy of the
excited state.
\end{abstract}
\vspace{1.0cm}

\section { INTRODUCTION}

Color transparency (CT) can be defined as the absence of or reduced final state
interaction of a nucleon with the nucleus in semiexclusive high momentum
transfer processes. Examples are the (e,e'p) and (p,pp) reactions occuring
on nuclear targets, in which the detected protons have enough energy to
ensure that no pions are produced.
For color transparency to be observed three
conditions have to be met \cite{fms94}.

(i) A small wave packet is formed in a high momentum transfer
reaction.
This wave packet is sometimes dubbed a point like
configurations(PLC).

(ii) A small wave packet interacts weakly with the nucleus.

(iii) The wave packet escapes the nucleus before expanding.

These conditions are true in the perturbative regime \cite{brodsky,mueller},
 but have to be
tested in the nonperturbative regime. Lattice QCD is a natural tool to
use in this investigation.
It can be used
to investigate the form of the hadron-hadron interaction \cite{green},
to probe the wave packet (hopefully PLC)
 produced in a high momentum transfer reaction,
and to examine the expansion of the PLC.

The separation of the interaction into the hard high momentum transfer part
and the soft final state interaction part is crucial for CT.
To be specific consider (e,e'p) reaction.
In this case the hard process is the absorption of the virtual photon by a
nucleon in the nucleus. An additional hard interaction is expected to be
suppressed by the experimental kinematics, and the final state interaction of
the PLC with the nucleus is of a soft low-momentum-transfer nature.
There are two amplitudes to be compared. The first one is for the
proton to escape the nucleus without interaction. The second one is for the
proton to be detected after the PLC is scattered by the soft interaction with
the nuclear medium. CT can be obtained if the ratio of the second amplitude
to the first one vanishes in the limit of $Q\rightarrow \infty$.

Ideally one would like to reconstruct on the lattice
the process described above: to create a wave packet by acting
with the electromagnetic current operator on the proton,
and to evaluate the amplitude of this wave packet to be transformed
 into a proton by some realistic
nucleon-PLC interaction. In this paper we will not take on this
challenging and complex problem.
 We concentrate on one aspect of the question - the
expansion of a model small size wave packet. To further simplify our task
we consider two-quark pion instead of three-quark proton.

 We start by defining the
necessary lattice constructions in Sec.2. A point source, one of the
conventionally used interpolating fields, is used to generate the small
size wave packet - PLC. A pion is detected in the final state after the PLC
experiences the soft interaction.
A model form of the soft
interaction  $\widehat{b}^2$ \cite{millerbsqr} is used, with $b$ representing
the transverse separation of the quarks. The wave packet undergoes
 Euclidean time evolution.
The quantity $b^2(t)$, which measures the strength of the soft interaction
of the evolving PLC with the nuclear medium,
is calculated as a function of the Euclidean time $t$.
The analytical continuation to the Minkowsky time $\tau$ is  then
performed. In Sec.3 the lattice details are given.
The purpose of  using lattice QCD is to obtain
matrix elements from the first principle calculations.
 Unfortunately it entails the Euclidean time evolution.
Our procedure is to treat the
wave packet as a coherent sum of physical states, so
the real problem is how to detect
as many states as possible before they decay away. These states are
extracted by a many pole fit of the lattice results. This procedure is
complicated and controversial. We discuss it for the two-point
correlation function in Sec.4.
To perform the analytical continuation the $b^2(t)$ is fit with the
sum of three exponentials.
The details of the $b^2(t)$ three pole fit are given in Sec.5.
 The results of the analytical continuation
and their discussion are presented in Sec.6. The relative merits of using
a full or a diagonal covariance matrix to fit observables is discussed in the
Appendix.

\section{  FORMALISM}

We want to consider the expansion of a small size wave packet.
To form such a wave packet we take advantage of the point
interpolating field $\phi({\bf x},t)$, which has the following form
in the pseudoscalar channel
\begin{equation}
\phi({\bf x},t) = d({\bf x},t) \gamma_5 \overline{u}({\bf x},t)
\end{equation}
By acting on the vacuum $\phi({\bf x},t)$ creates an object similar to a
PLC, since the quark and the antiquark are created at the same site.
This wave packet can be expanded in a complete set of zero momentum
states $|\nu>$:
\begin{equation}
\sum_{{\bf x}} \phi({\bf x},0) |0> \hspace{0.1in}=>
\hspace{0.1in} \sum_\nu(\int d\nu) c_\nu |\nu>\equiv|PLC>   \label{eq:PLC}
\end{equation}
where the summation over ${\bf x}$ projects out the zero momentum states.
The coefficient $c_\nu$,
\begin{equation}
c_\nu=< \nu| \phi(0,0) |0>/\sqrt{2E_\nu},
\end{equation}
is the strength with which the state $|\nu>$ is excited
from the vacuum, $\sqrt{2E_\nu}$ is a normalization factor.
We have to note that even though the interpolating field creates a
quark and an antiquark at the same site, there is a non-zero probability
to detect them separated by some distance at the same time they are
created. This occurs because fermion propagators do not
vanish outside of the light cone.

If we transport the $|PLC>$ with the Euclidean transfer matrix
$\exp(-\widehat{H}t)$ and then contract it with itself we obtain the
two-point correlation function $G_2(t)$:

\begin{eqnarray}
G_2(t)= \sum_{{\bf x}}< 0| \phi({\bf x},t) \phi(0,0) |0>=
\nonumber\\
=  \sum_\nu(\int d\nu){|c_\nu|^2 e^{-E_\nu t}}   \label{eq:G2}
\end{eqnarray}
$G_2(t)$ plays a complementary role in our investigation, as will be seen
below.

Our main objective is to evaluate the strength of the soft interaction
of the expanding PLC with the nucleus. This interaction
can be described (\cite{millerbsqr} and references therein) by the
operator $\widehat{b}^2$ with $b$ representing the
transverse separation of the quarks in the PLC
\footnote{For the real PLC the longitudinal direction is given by the momentum
transfer in the hard interaction.
Unlike real PLC our wave packet is spherically symmetrical so we
arbitrarily choose the $z-$ direction to be longitudinal.}:
\begin{equation}
b^2(t)=\frac{1}{c_1}<\pi|e^{\hat{H}t} \widehat{b^2} e^{-\hat{H}t} |PLC>.
\label{eq:soft}
\end{equation}
Here we normalize it by $c_1=<\pi|PLC>$,
 which represents the amplitude of the
detecting the pion without the soft interaction.
To find the matrix element (~\ref{eq:soft})
we need to know the wave functions of
the PLC and the pion. There is no unique way to define the wave functions
on the lattice. We use the gauge invariant formulation of
equal time Bethe-Salpeter amplitude  $\Psi({\bf r},t)$ \cite{negele1}
\begin{eqnarray}
\Psi ({\bf r},t) = \sum_{{\bf x}}< 0| \overline{d}({\bf x},t)
\gamma_5 U({\bf x\rightarrow  x+r},t) u({\bf x+r},t)  \phi(0,0) |0>.
\end{eqnarray}

The Bethe-Salpeter amplitude in the gauge invariant formulation is
known \cite{negele1} to underestimate the spatial extent of the pion
wave function. But it does not  suffer from finite lattice size errors
and is hoped to give a qualitatively correct estimate of the effects
under consideration.

In a fashion similar to $G_2(t)$
$\Psi ({\bf r},t)$ can be expanded in a complete set of states
\begin{equation}
\Psi ({\bf r},t) =\sum_\nu(\int d\nu) c_\nu \psi_\nu({\bf r}) e^{-E_\nu t},
\end{equation}
where
\begin{equation}
\psi_\nu({\bf r})=<0|\overline{d}(0,0) \gamma_5 U({\bf r},t)u({\bf r},t)
|\nu>/\sqrt{2E_\nu}
 \end{equation}
is the Bethe-Salpeter amplitude of the state $|\nu>$.

$\Psi ({\bf r},t)$ is measured on the lattice and $b^2(t)$ is calculated
according to
\begin{equation}
b^2(t)=\int d^3{\bf r} \Psi({\bf r},t)
\psi_{\pi}({\bf r},t)  (r cos \theta )^2.
\end{equation}
The ground state wave function $\psi_{\pi}({\bf r},t)$ is obtained from
$\Psi ({\bf r},t)$ in the limit $t\rightarrow \infty$.

$b^2(t)$ has the following spectral representation
\begin{equation}
b^2(t)=  \sum_\nu(\int d\nu) \frac{c_\nu}{c_1}
 b^2_\nu e^{-(E_\nu-E_{\pi}) t},
   \label{eq:b^2spectr}
\end {equation}
where
\begin{equation}
b^2_\nu=\int d^3{\bf r} \psi_\nu({\bf r},t)
\psi_{\pi}({\bf r})  (r cos \theta )^2.
\end{equation}

The spectral representation (~\ref{eq:b^2spectr}) can be used
to perform the  transition from the Euclidean
time to the Minkowsky time. If we knew the $E_\nu$'s, $c_\nu$'s, and
$b_\nu$'s for all the contributing states, then a mere substitution
 $t\rightarrow i\tau$ would give us $b^2(\tau)$, which describes the
expansion of the PLC in the real time.
This is certainly too much to ask for.
Since we work with the Euclidean time most of the states decay away too
quickly to be detected.

We assume that the spectral density
can be represented as a sum of several sharp poles. With this assumption
the goal is to try to determine the $E_\nu$'s, $c_\nu$'s, and
$b_\nu$'s for as many states as possible.
We perform the many pole fit of $b^2(t)$ with the fit function $f_b(t)$:
\begin{equation}
f_b(t)=\sum_n^{N_p}a_n e^{-E_n t},                \label{eq:f}
\end{equation}
where $N_p$ is the number of the poles.
 However, there are several sticky points
here. First, there has been some concern expressed in the literature
\cite{derek} about the validity of such a many pole fit.
Second, it is not {\em a priori} obvious that those few states we are able
to extract will be enough to form a small or at least reduced size wave
packet.
In Sec.~\ref{fit} and Sec.~\ref{b^2fit}
we show that a simultaneous three pole fit of $b^2(t)$ and $G_2(t)$
can be  reliably performed to yield the parameters for the three lowest states.
These three states
do form a reduced size wave packet, whose expansion is considered
in Sec.~\ref{expansion}
 But first we give a summary of
the lattice calculations in the next section.

\section{LATTICE DETAILS}
The calculations are performed for $SU(2)$ gauge theory with the
coupling constant $\beta=2.5$. $SU(2)$ is chosen to increase the
efficiency of the calculation and improve statistics. The size of the
lattice is $12^3\times 24$. 60 quenched gauge field configurations
are generated
using Metropolis method with overrelaxation \cite{overrelax}.
 The first configuration
is selected after 2000 thermalization sweeps and all the consecutive ones
after 1000 sweeps.

The Wilson propagators are calculated  for three values of the hopping
parameter $\kappa=0.146, 0.148, 0.149$.
Periodic boundary condition in the
spatial directions and Dirichlet boundary conditions in the time
direction are used. A standard procedure \cite{negele1}
is used to construct the
two-point correlation function and the Bathe-Salpeter amplitude out of
the quark propagators.

The lattice spacing is determined from the $\rho$ mass \cite{romass}
to be $a=0.09\pm0.012$ fm, with is in quantitive agreement with the
results \cite{green} obtained from the string tension.

The quark mass $m_q$ is conventionally associated with the hopping
parameter $\kappa$: $m_q=(1/2\kappa-1/2\kappa_{cr})a^{-1}$. Extrapolation
to the limit $m_{\pi}=0$ yields $\kappa_{cr}=0.151(1)$. With this $\kappa_{cr}$
we obtain three values of the quark mass $m_q= 275, 174, 124 $ Mev.

The covariance matrix fit is performed to determine the parameters needed.
A detailed description of this procedure is given in Sec.~\ref{fit}.
All statistical errors are estimated by the single elimination
jacknife \cite{jacknife}.
 The point source is placed at the time slice $t=5$
and the fitting is performed over the range $t=6$ through $t=20$.

\section{ MANY POLE FIT}   \label{fit}

A many pole fit involves many problems,
 some of which are mentioned in \cite{derek,kilcup}.
To address these problems we perform a detailed analysis of the fit
of the two-point correlation function $G_2(t)$.
The fit function $f_G(t)$
\begin{equation}
f_G(t)=\sum_{n=1}^{N_p} c_n e^{-E_n t}
\end{equation}
 corresponds to $N_p$-pole fit.

We would like to emphasize that our goal here is to extract as many states
as possible to be able to form a small size wave packet.
This is contrary to the intention of the majority of the papers in the
field, where the isolation of one state, usually the ground state, is pursued.

A standard approach in determining
 parameters $c_n$'s and $E_n$'s is to minimize
the goodness-of-fit statistic $\chi^2$ .
\begin{equation}
\chi^2 =  \sum_{t,t'} (G_2(t)-f_G(t))C(t,t')^{-1}(G_2(t')-f_G(t')
\end{equation}
The jacknife method is used to calculate the covariance matrix $C(t,t')$
\begin{eqnarray}
C(t,t')= \frac {1}{N_c-1}  \sum_{k=1}^{N_c}
(G_2^{[k]}(t)-G_2(t))
(G_2^{[k]}(t')-G_2(t'),
\end{eqnarray}
where  $N_c$ is the number of configurations, $G_2^{[k]}(t)$ are found
for a subset of
$N_c -1$ configurations with $k'th$ configurations omitted.

 Simulated annealing is our choice of the minimization technique.
The full covariance matrix is used in the minimization(correlated $\chi^2$
fit).
There have been studies \cite{michael,seibert}
to show that for a small data sample use of
the full covariance matrix is unreliable.
In \cite{kilcup} a use of several well defined eigenmodes of the covariant
matrix is proposed. The use of a full covariant
matrix versus the diagonal part thereof poses an interesting problem.
We address this problem in the Appendix, where we consider a simple example
 amenable to the analytical treatment. The conclusion is reached
 that a correlated fit gives
better results if the correlations are appreciable. This is generally the
case for lattice calculations, and it is true for our calculations.
 Moreover, we consider 60 configurations to be sufficient for
the maximum of 9 degrees of freedom we have while fitting $G_2(t)$.

The fitting is performed over a time range extending
from $t_{first}$ to $t_{last}$. We choose $t_{last}=20$ to
exclude the boundary effects and $t_{first}$ is varied from 6 to 14.
The number of states used in a fit is determined by the $t_{first}$.
One of the problems with a many pole fit \cite{derek}
 is that the result depends on the $t_{first}$ used for the fit, and it is
not {\em a priori} clear how many states should be included in a fit.

The strategy developed to tackle this problem is as follows.
 We start with $t_{first}$ for which
only the ground state contribution is significant, and one-pole fit is
performed. As $t_{first}$ decreases more states with higher masses come into
play. To decide between two fits with different number of poles
for a particular $t_{first}$ the preference is given to a
fit with smaller value of $\chi^2$ per degree of freedom.
The results for the $E_n$'s for the hopping parameter $\kappa=0.149$
are shown in Fig.~1. For $t_{first}=12 \div 14$ one state is
enough, for $t_{first}=9 \div 11$ two states had to be included,
and for $t_{first}=6 \div 8$ the $G_2(t)$ is best fit with three states.
The results for the other two hopping parameters are similar.
The values of the $\chi^2$ per degree of freedom
  show that the fit for the longest time range
is the best( Table 1).

The pattern in the results for $E_n$ is obvious. The statistical error
is the greatest when a new state is included in the fit,
 it then decreases and jumps back, when another state is included.
To explain this pattern the following plot is made.
The individual contributions $a_n e^{-E_n t}$ of the states to the two-point
correlation function $G_2(t)$ are shown in Fig.~2. They are compared with
the statistical error on the $G_2(t)$.
Two time slices
$t_{first}=12$ and $t_{first}=9$ are important, since at those times
the second and third states, go above the statistical noise.
Those are precisely the time slices for which the contribution of these states
is included in the fit.

 A simple rule follows: a state should be included
 if its contribution is greater than the statistical
error; it should not be included if its contribution is
less than the statistical error.
The best results of the fitting are obtained for
the longest time span of the highest energy state used in the fitting.
An interesting consequence is that, in general, the longest time range is not
necessarily the best for determining the parameters.
Consider the following example. Suppose that in Fig.~1
the results of the three pole fit started at $t_{first}=8$ (i.e. the
source was at $t=7$ and the results for $t_{first}=6$ and $7$ were not
present).
In that case
a reliable determination of the
third state's parameters would be impossible; furthermore
we would be forced to use the shorter
time range with $t_{first}=9$
to determine the
parameters of the two lowest states.

Another interesting observation concerns the fact that $\chi^2$ has many
minima in the multi-dimensional space of $c_n$'s and $E_n$'s. The ``real"
minimum is not always the global one. By the ``real'' we mean the one
we believe to be the best fit. In our case the best fit is obtained for
the $t_{first}=6$. We ran the minimization for $t_{first}=7$ and
$t_{first}=8$ and found local minima very close to the ``real''
one. The gain in $\chi^2$ compared to the global minimum
is only  $\approx 0.1$ per degree of freedom.

To support this argument the following simulation is performed.
We work our way backwards. Four sets of  $\tilde{c}_n$ and $\tilde{E}_n$
are chosen.
A ``correlation function'' $\tilde{G}(t)$ is calculated:
\begin{equation}
\tilde{G}(t)=\sum_{n=1,4}\tilde{c}_n\exp(-\tilde{E}_n t).
\end{equation}
Then statistical noise is added and 60 ``configurations'' are generated.
The subsequent fitting procedure yield results that have all the basic
features of the fitting of the lattice data(Fig.~3).
The input values of the parameters of three states
 are recovered (within the error bars). The fourth state is intentionally
chosen to decay below the statistical noise by $t=6$.
In this simulation we know what the correct values of the parameters are,
and we can verify that for $t_{first}=7$ and
$t_{first}=8$ there are local minima very close to these correct values.

We conclude that for the given choice of interpolating field and lattice
spacing three pole fit of the two-point correlation function can be
performed and the values of the excitation strength $c_n$ and the mass
$E_n$ reliably determined. These values are presented in Table 2.

\section{$b^2(t)$ MANY POLE FIT}     \label{b^2fit}

The determination of the $c_n$'s and $E_n$'s is interesting, but it is does not
constitute the main objective of this paper.
What does is the strength of the soft interaction $b^2(t)$, which
we fit with the Eq.(~\ref{eq:f}).
 As was indicated in the introduction, a fit of $b^2(t)$ alone does not
give a satisfactory result. To understand it, note that
  a coefficient $a_n$ in Eq.(~\ref{eq:f})
is proportional to the excitation strength $c_n$ and the soft interaction
transition matrix element $b^2_n$ of the state $|n>$.
For the  three states, that we are able to extract from the two-point
correlation function $G_2(t)$, the excitation
strength $c_n$ grows with $n$. (Physically it means that their wave
functions have a better overlap with the
wave function of the initially formed wave packet.)
Unlike the coefficients $c_n$ $b^2_n$ turned out to decrease rapidly
with $n$. As a consequence the contribution to the $b^2(t)$ of
the exited states disappears below the statistical noise much sooner
than for the $G_2(t)$. This forces us to fit  $b^2(t)$ together with
$G_2(t)$.

 We would like to note here that the procedure we use to compute
$b^2(t)$ is akin to the smearing technique
used in \cite{smear1,smear2} to isolate the contribution of meson and baryon
 ground states. However, let us stress it again, this is not our intention.
This is an ``accident'', but an ``accident'' with some physical consequences,
 which are discussed below.

The simultaneous fit of $b^2(t)$ and $G_2(t)$ is performed
for the longest time range possible ($t_{first}=6$). The results are shown
in the Table 3.
 $b^2_n$ goes down by a factor of 2 with every excited state.
Another important feature is that $b^2_3$ is negative. This allows for the
formation of a reduced size wave packet.

Let's discuss the physical meaning of these results. Do they have any relevance
for color transparency? Again consider (e,e'p) process. In this
case the coefficients $c_\nu$ defining the PLC are the elastic and
inelastic form factors
of the ground state. In general we can expect these coefficients to exibit
a resonant behavior - to have a peak for a certain excited state.
 But the soft interaction supresses the contribution of the higher
excited states, as $b_\nu$ decreases rapidly with $\nu$ \cite{our}.
 This suppression
 plays a crucial role in achieving a slow expansion rate of the PLC.
 Our result turned out to have these basic features. The $c_n$ grows with
$n$, whereas $b_n$ decreases rapidly.
The former is only a result of our choice of the model wave packet
and in this sense is arbitrary.
 The latter, however,  reflects
the properties of hadronic states in lattice QCD, and to the extent that
we believe lattice QCD, this is an indication of the possibility of the
suppression of the contribution of the higher excited states by the soft
interaction.

\section{MINKOWSKY TIME EXPANSION}   \label{expansion}

Having obtained the parameters $E_n,c_n,b^2_n$ for the three lowest states,
 we substitute $t\rightarrow i\tau$ in Eq.(~\ref{eq:b^2spectr})
and plot the real part of $b^2(\tau)$ (Fig.~4).
\begin{equation}
Re(b^2(\tau))= \sum_{n=1}^3 \frac{c_n}{c_1} b^2_n cos(E_n-E_1)\tau
\end{equation}
The band shown in Fig.~4 corresponds to the 60 sets of 59 configurations
used in the jacknife evaluation. Each of the curves shows an expansion.
We do not compute a set of curves simply using the errors quoted
in Table 2 and Table 3. Such a procedure would incorrectly ignore the
correlations between the parameters $E_n$, $c_n$, and $b^2_n$.

A look at Fig.~4 shows us that
the wave packet under investigation is not
really small. According to Fig.~4, $b^2(0)\approx 3/4 ~b^2_1$,
where $b^2_1=<\pi|b^2|\pi>$ is the ground state contibution.
 (In the Fig.~4 $b^2_1$ is represented by the solid horizontal line.)
This is because we could recover only three out of the many states
comprising the wave packet at the source $t=5$.
Furthermore, even at  $t=5$
 the wave packet is not quite point like. As was mentioned above
, there is a non-zero probability for the quark and the antiquark to be
separated by some distance at the time they are created by the point
interpolating field.

These features make it difficult to define one number $\tau_{exp}$
to typify the time for the wave packet to expand.
One way is to define $\tau_{exp}$ as the time it takes for
$b^2(\tau)$ to reach the ground state contribution $b^2_1$.
 This expansion time is found to be $\approx0.03$ fm, which
is certainly much smaller than the root mean square radius of the pion
$\approx 0.4$ fm here
\footnote{ This small number is the consequence of using
the gauge invariant formulation of the Bethe-Salpeter wave function.}.
However, this estimate is misleading, since the initial wave packet is
not small. The initial wave packet size is determined by the
coefficients $c_n$ as well
 as $b_n^2$. Here the set $c_n$ determines the spectral density of
the operator $\phi(x)$. These numbers are only related to color transparency
physics if they also represent the
elastic and inelastic form factors. Thus we are free to consider  the
consequences of using different values of $c_n$. Varying $c_n$
 we can obtain an upper limit for the expansion time for the given
energies $E_n$ and soft interaction matrix elements $b_n^2$. This is achieved
for $c_3=0$ and $c_2<0$ and equal to $\tau_{exp}\approx 2$ fm.

The results for the other two hopping parameters are similar to
the one shown in Fig.~4, but with bigger statistical error.
For the purpose of our investigation we are not interested in
taking the  physical limit $E_1\rightarrow m_{\pi}$.
The point is that the expansion time is determined by the energy
differences $E_n-E_1$ as well as $c_n$ and $b^2_n$.
 The coefficients $c_n$ and $b^2_n$ are virtually independent
of the quark mass. The lowest value of the
 energy gap for a pion is much larger, than for a proton.
 By taking the physical limit to $m_{\pi}\approx0$ we only
increase the mass gap and decrease the expansion time.

There is one more point to be made. The subject of our investigation
 is the expansion of a wave packet in its rest frame.
 If color transparency is to be observed, it will be
 in a high momentum transfer reaction,
 when the escaping wave packet is moving with a speed close to the
speed of light. Time dilation
will increase the expansion time. In fact, color transparency can
be achieved for any small size wave packet,
consisting of a finite number of states,
 as long as the momentum transfer is much greater than
the energy of the states. A real PLC produced in a high momentum transfer
reaction consists of an infinite number of
states. How the energy of the important states changes with the momentum
transfer will decide the fate of color transparency.

\section{CONCLUSION}
The expansion of a reduced size wave packet on the lattice is investigated.
The wave packet is created with the pseudoscalar point interpolating
field.
 The quantity representing the strength of the soft interaction of the
expanding wave packet with the nucleus is measured as a function of
the Euclidean time.  We treat the expanding wave packet as a
coherent sum of physical states.
Most of the states decay away before we can detect them.
The three lowest energy states we are able to recover allow for the
formation of a reduced size wave packet. With the parameters
of these states we are able to do the transition to the Minkowsky time.
 A Minkowsky time expansion picture is obtained.

What is the importance of our work and where do we go from here?
 For the first time three states - a ground state and two excited states-
 are obtained from the lattice calculations.
(One may attempt to recover more excited states by using a smaller
lattice spacing $a$.)
The matrix elements of
the soft interaction between the ground state and the two excited states
are important for color transparency. These are found to decay rapidly with
 energy. This feature is an important component of color transparency.
(It would be useful to verify that similar results are obtained for baryons.)
Furthermore, the present results show that
a lattice evaluation of wave packet expansion is possible.

In the present results, the initial wave packet is not small, and its
relation
to color transparency physics is unclear. Our next step
will be to attack this problem by
studying the wave packet (hopefully a PLC) created
by the electromagnetic current operator.

\vspace{0.3in}
It is our pleasure to thank Steven Sharpe for useful discussions and
encouragement. This work was supported in part by the U.S. Department
of Energy. The lattice calculations were performed on the
UW Nuclear Theory DECstation 3000-600 AXP.

\appendix
\section{FULL VS. DIAGONAL COVARIANCE MATRIX FIT}

There have been several papers, in which the attention has been
brought to the fact
that the use of the full covariance matrix for the correlated data is
unreliable for small data samples \cite{michael,seibert}.
We use a simple example to show analytically that there are instances,
when the use of the diagonal part of a covariance matrix to fit data
(uncorrelated $\chi^2$ fit) with a
theoretical expression gives a better result than the full covariance
matrix (correlated $\chi^2$ fit),
even in the limit of infinite number of samples.
Note that when we use the diagonal part, we fit {\em correlated} data
and simply ignore the correlations among them.

Consider fitting results of the $N_{sample}$ measurements
 of $N$ quantities $x$ by a constant I.
We assume that all $\overline{x_i}$'s
\begin{equation}
\overline{x_i}=\frac{1}{N_{sample}}\sum_{\alpha=1}^{N_{sample}} x^{\alpha}_i
\end{equation}
do reach the same constant value $J$ in the limit
$N_{sample}\rightarrow \infty$ and that the
correct infinite $N_{sample}$ form of the covariance matrix $C_{ik}$ is known.

Then we can minimize $\chi^2$,
\begin{equation}
\chi^2=\sum_{ik}(\overline{x_i}-I) C_{ik} (\overline{x_k}-I),
\end{equation}
with respect to $I$ and find $I$ in two cases:  $I_{full}$, when
the full covariance matrix $C$ is used, and $I_{diag}$, when only the diagonal
elements of the covariance matrix $C$ are used.
\begin{eqnarray}
I_{full}= \frac{\sum_{ik} C_{ik}^{-1} \overline{x_k}}{\sum_{ik} C_{ik}^{-1}},
\hspace{0.6in}
I_{diag}=\frac{\sum_{i} \overline{x_i}/C_{ii}}{\sum_{i}1/C_{ii}}.
\end{eqnarray}

Both $I_{full}$ and $I_{diag}$ are unbiased estimators of the true value $J$,
since in the limit $N_{sample}\rightarrow \infty$ they both are equal to $J$.
The preference should be given to the estimator with the smaller standard
deviation. Results for $I_{full}$ and $I_{diag}$ will vary from one set of
 $N_{sample}$ samples to another with the variance
$C_{I}=\overline{I^2}-(\overline{I})^2$ equal to
\begin{eqnarray}
C_{I_{full}}=(\sum_{i k} C^{-1}_{ik})^{-1}  \hspace{0.6in}
C_{I_{diag}}=(\sum_{i}\frac{1}{C_{ii}})^{-1}.
\end{eqnarray}
{}From this general form it is clear that we can not expect
$C_{I_{full}} < C_{I_{diag}}$ to be always true.
Consider a situation, when the diagonal elements of the matrix $C$ are much
greater than nondiagonal:
\begin{equation}
\frac{C_{ik}}{C_{ii}} \ll 1; \hspace{0.5in}
\frac{C_{kk} C_{ik}^2}{C_{ii}} \ll 1; \hspace{0.1in} i\neq k,
\end{equation}
and we can expand $C^{-1}$ to get for $C_{I_{full}}$
\begin{equation}
C_{I_{full}}=(\sum_{i}\frac{1}{C_{ii}}-
\sum_{i\neq k}\frac{C_{ii}}{C_{kk}} C_{ik})^{-1}
\end{equation}
 The uncorrelated fit yields better results in the case of
 prevailing positive correlations, the correlated fit is better
in the case of  prevailing negative correlations.
 But positive correlations do not necessarily
render a correlated fit worse than uncorrelated one.
Consider a simple example of $N=2$, when the expressions for
$C_{I_{full}}$ and $C_{I_{diag}}$ become transparent:
\begin{eqnarray}
C_{I_{full}}=\frac{C_{11}C_{22}-C_{12}^2}{C_{11}+C_{22}-2C_{12}}
\hspace{0.6in}
C_{I_{diag}}=\frac{C_{11}C_{22}}{C_{11}+C_{22}}
\end{eqnarray}
Again, as in the case of small correlations, a correlated fit is better
for negative correlations $C_{12}<0$. But there is a region of big positive
 correlations
\begin{equation}
C_{12} > \frac{2 C_{11}C_{22}}{C_{11}+C_{22}}
\end{equation}
where a correlated fit gives better results.

Another issue is using $\chi^2$ to estimate how good a fit is.
For the correlated fit the infinite $N_{sample}$ limit of $\chi^2_{full}$
is $N-1$, which is exactly 1 per degree of freedom.
In the case of the uncorrelated fit $\chi^2$ reaches this limit only if the
data is genuinely uncorrelated $(C_{ik}=0,  i\neq k)$.
When we ignore the correlations
and use the diagonal part of the correlation matrix to obtain
the result for $I$, they pop up in the expression for  $\chi^2_{diag}$:
\begin{equation}
\chi^2_{diag}= N-1 -\sum_{i\neq k} \frac{C_{ik}}{C_{ii}C_{kk}}
/\sum_i (C_{ii})^{-1}
\end{equation}
One therefore can not always use $\chi^2$ to evaluate an uncorrelated fit.

\begin{table}
\begin{tabular}{cccccccccc}\hline
\\
$t_{first}$ &\nh 6 &\nh 7 &\nh 8 &\nh 9 &\nh 10 &\nh 11 &\nh 12 &\nh 13
&\nh 14   \\
\\ \hline \\
$\chi^2$ &\nh 1 &\nh 1.2 &\nh 1.6 &\nh 1.5 &\nh 1.8 &\nh 2 &\nh 2.3 &\nh 1.6
 &\nh 0.3 \\
\\
\hline
\end{tabular}
\caption{$\chi^2$ per degree of freedom for different time ranges}
\end{table}
\begin{table}
\begin{tabular}{llll} \hline\hline
\\
Parameter & \h $\kappa_1=0.146$ & \h $\kappa_2=0.148$ & \h $\kappa_3=0.149$ \\
\\
 \hline
\\
$E_1 a$ & \h  0.49(1)      & \h   0.39(4)     & \h   0.33(5)        \\
\\
$E_2 a$ & \h   1.2(1)       & \h   1.1(2)      & \h   1.0(2)         \\
\\
$E_3 a$ & \h   2.2(1)       & \h   2.2(2)      & \h   2.1(2)         \\
\\
$c_1$  & \h  0.39(3)      & \h   0.36(6)     & \h   0.35(6)        \\
\\
$c_2$  & \h  1.1(2)       & \h   1.1(2)      & \h   1.0(3)        \\
\\
$c_3$  & \h  2.3(1)       & \h   2.4(1)      & \h   2.4(6)  \\ \\ \hline \hline
\end{tabular}
\caption{ Energy $E_n$ and excitation strength $c_n$ of the three pole fit
of the two-point correlation function $G_2(t)$}
\end{table}

\begin{table}
\begin{tabular}{llll} \hline\hline
\\
Parameter & \h $\kappa_1=0.146$ & \h $\kappa_2=0.148$ & \h $\kappa_3=0.149$ \\
\\
 \hline  \\
$b_1^2$& \h  0.08(2)  & \h   0.08(2)     & \h   0.08(2)        \\
\\
$b_2^2$& \h  0.04(1)  & \h   0.05(2)     & \h   0.05(2)        \\
\\
$b_3^2$& \h  -0.02(1)  & \h   -0.02(1)   & \h   -0.015(13)         \\
\\
\hline \hline
\end{tabular}
\caption{ The soft interaction matrix elements $b_n=<n|b^2|1>$.}
\end{table}

\begin{figure}[p]
\vspace*{5in}
\hspace*{-2in}
\caption{The energies of three states as a function of the first
time slice $t_{first}$ used in the fitting. $t_{last}=20$ for all the fits.
The dash line shows the result for the longest time range.
 Hopping parameter $\kappa=0.149$.}
\end{figure}

\begin{figure}[p]
\vspace*{4in}
\hspace*{-2in}
\caption{Individual contribution of the three states to the two-point
correlation function to be compared to the statistical error of the
two-point correlation function $G_2(t)$. Hopping parameter $\kappa=0.149$.}
\end{figure}

\begin{figure}[p]
\vspace*{5in}
\hspace*{-2in}
\caption{The energies of three states for the simulated $\tilde{G}(t)$.
 $t_{last}=20$ for all the fits.
The dash lines show the correct input values of the energies.}
\end{figure}

\begin{figure}[p]
\vspace*{5in}
\hspace*{-2in}
\caption{Expansion of the wave packet in real time. The solid curve
corresponds to the average over all 60 configurations. The dash curves
are obtained for 60 sets of 59 configurations that were used in the
jacknife evaluation. The horisontal solid line is the ground state
contribution $b^2_1$. Hopping parameter $\beta=0.146$.}
\end{figure}

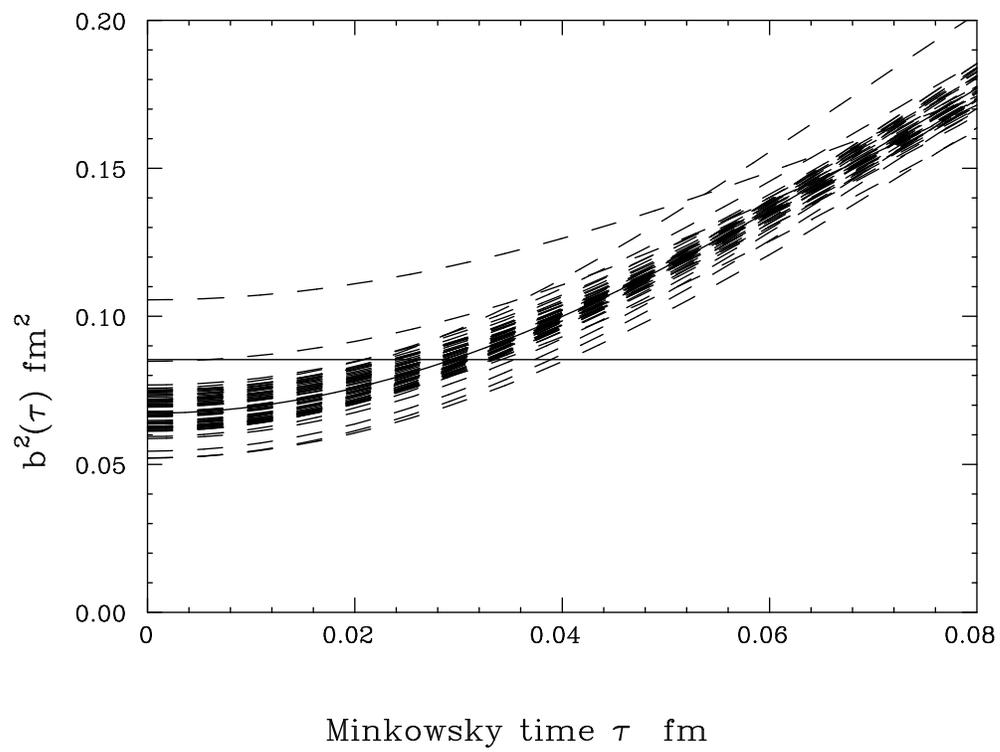
\end{document}